\documentclass[%
reprint,
nofootinbib,
amsmath,amssymb,
aps,
prd
floatfix,
showkeys
]{revtex4-1}

\usepackage{caption}
\usepackage{subcaption}

\usepackage{graphicx}  
\usepackage{amsmath}   
\usepackage{amssymb}   
\usepackage{bm} 
\usepackage{dcolumn}
\usepackage{color}
\usepackage{mathrsfs}
\usepackage{amsfonts}
\usepackage{varioref}
\usepackage{mathrsfs}
\usepackage{graphicx}
\usepackage{latexsym}
\usepackage{amsmath}
\usepackage{amssymb}
\usepackage{textcomp}
\usepackage{amsbsy}
\usepackage{graphics}
\usepackage{epstopdf}
\usepackage{color}
\usepackage{enumitem}
\usepackage{array,multirow} 
\usepackage{float}
\usepackage{extarrows}
\usepackage{url}
\usepackage{caption}

\allowdisplaybreaks

\input epsf

\usepackage{hyperref}
\hypersetup{
    colorlinks=true,
    citecolor=blue,
    linkcolor=blue,
    filecolor=magenta,      
    urlcolor=magenta,
}

\newcommand{\Omegam}{\Omega_\text{m}}
\newcommand{\Omegar}{\Omega_\text{r}}
\newcommand{\Omeganu}{\Omega_\nu}
\newcommand{\Omegab}{\Omega_\text{b}}
\newcommand{\Omegac}{\Omega_\text{c}}

\newcommand{\omegab}{\omega_\text{b}}
\newcommand{\omegabc}{\omega_\text{bc}}

\newcommand{\Neff}{N_\text{eff}}

\begin{document}

\tolerance=5000

\title{Exponential $f(R)$ cosmology with massive neutrinos as a dynamical dark energy framework}
\author{Simone~D'Onofrio$^{1}$\, \thanks{donofrio@ice.csic.es}}
\author{Sergei~Odintsov$^{1,2}$\,\thanks{odintsov@ieec.uab.es}} 
\author{Tiziano~Schiavone$^{3,4,1}$ \, \thanks{tiziano.schiavone@sissa.it}}
\affiliation{
$^{1)}$ Institute of Space Sciences (ICE, CSIC) C. Can Magrans s/n, 08193 Barcelona, Spain}
\affiliation{$^{2)}$ ICREA, Passeig Luis Companys, 23, 08010 Barcelona, Spain}
\affiliation{$^{3)}$ SISSA-International School for Advanced Studies, Via Bonomea 265, 34136 Trieste, Italy}
\affiliation{$^{4)}$ INFN, Sezione di Trieste, Via Valerio 2, I-34127 Trieste, Italy}


\tolerance=5000

\begin{abstract}
The exponential $f(R)$ gravity model provides a theoretically well-motivated extension of General Relativity, introducing a modified gravitational dynamics at late times consistent with a dynamical dark energy scenario, while recovering the $\Lambda$CDM-like regime at high redshifts with a smooth transition. Using a Bayesian Markov Chain Monte Carlo (MCMC) analysis, we constrain the parameters of the exponential $f(R)$ model in combination with the total neutrino mass $\sum m_\nu$, employing the latest measurements from cosmic chronometers, the DESI DR2 BAO data, the CMB acoustic scale, and the Pantheon+ supernovae compilation, comparing the results with the $\Lambda$CDM and the $w_0w_a$CDM models. Our results show that the exponential $f(R)$ model remains consistent with current observations while partially alleviating the Hubble tension and the neutrino mass problem relative to $\Lambda$CDM, although the constraints on $\sum m_\nu$ are tighter than those obtained for the phenomenological $w_0w_a$CDM scenario. These results indicate that the interplay between modified gravity and neutrino physics in the late Universe may offer a viable framework for further investigating cosmological tensions.
\end{abstract}

\maketitle

\section{Introduction}
The standard cosmological framework, the $\Lambda$ cold dark matter ($\Lambda$CDM) model, is based on General Relativity (GR) as the underlying theory of gravity and provides a highly successful and self-consistent description of cosmic evolution from the early Universe to the present epoch. Supported by a wide range of observations at galactic and cosmological scales, including probes such as the cosmic microwave background (CMB), baryon acoustic oscillations (BAO), and Type Ia supernovae (SNe Ia), the model accurately reproduces key features of the Universe, such as recombination, structure formation, and the current accelerated expansion.

Despite its remarkable success, $\Lambda$CDM faces persistent theoretical and observational challenges. The cosmological constant $\Lambda$, interpreted as a vacuum energy with negative pressure, is required to explain late-time acceleration \cite{Riess_1998,Perlmutter_1999}, while cold dark matter accounts for galactic rotation curves, gravitational lensing, and large-scale structure formation \cite{Trimble:1987ee}. However, the physical nature of both components remains unknown. In particular, $\Lambda$ suffers from the fine-tuning and coincidence problems \cite{Weinberg:1988cp,Peebles:2002gy}, and no evidence for a dark matter particle beyond the Standard Model has yet been found.

Additionally, several observational discrepancies have recently emerged. The most significant issue is the so-called Hubble constant tension \cite{DiValentino:2020zio,DiValentino:2021izs,Bamba:2012cp,Abdalla:2022yfr,Perivolaropoulos:2021jda,CosmoVerse:2025txj}, which concerns the measurement of the current expansion rate of the Universe, i.e., the Hubble constant $H_{0}$. Local distance-ladder measurements by the SH0ES collaboration yield $H_{0}=\left(73.04\pm1.04\right)\,\textrm{km s}^{-1}\,\textrm{Mpc}^{-1}$ \cite{Riess_2022,Scolnic:2021amr,Brout_2022}, while early-Universe measurements inferred from CMB data within the $\Lambda$CDM framework give $H_{0}=\left(67.36\pm0.54\right)\,\textrm{km s}^{-1}\,\textrm{Mpc}^{-1}$ \cite{Planck_2020}. The $\sim5\sigma$ discrepancy between these values may indicate unaccounted systematic effects or, more intriguingly, the need for new physics beyond the standard cosmological model \cite{Vagnozzi:2019ezj,Capozziello:2025lor, Chaudhary:2025bfs,Chaudhary:2025uzr}. 


Furthermore, the recent BAO measurements from the Dark Energy Spectroscopic Instrument (DESI) collaboration, based on galaxies, quasars, and Lyman-$\alpha$ forest data, indicate a deviation from the $\Lambda$CDM model at a confidence level of about $2.8-4.2\,\sigma$ \cite{DESI:2025zgx}, depending on the SNe Ia dataset used in the joint analysis. This deviation has been interpreted as evidence for a dynamical dark energy component, consistent with the Chevallier–Polarski–Linder parameterization (CPL), or the so-called $w_{0}w_{a}$CDM model \cite{Chevallier:2000qy,Linder:2002et}. However, such a parameterization remains phenomenological for a slowly running cosmological constant and lacks a solid theoretical foundation (see also \cite{Giare:2024gpk} for the analyses of alternative parameterizations).

These theoretical and observational challenges have motivated the exploration of alternative or extended cosmological frameworks beyond $\Lambda$CDM. A particularly compelling approach involves modifying the gravitational sector itself. Within this context, $f(R)$ gravity represents one of the most widely studied and mathematically tractable extensions of GR \cite{Buchdahl:1970ynr,Song:2006ej,Sotiriou:2008rp,Nojiri:2010wj,Capozziello:2011et,Nojiri:2017ncd}. By promoting the gravitational Lagrangian to a general function of the Ricci scalar, $f(R)$, these theories introduce an additional scalar degree of freedom that can drive cosmic acceleration without requiring an explicit cosmological constant. Several viable $f(R)$ models have been proposed that reproduce the observed expansion history and satisfy local gravity constraints \cite{Hu:2007nk,Starobinsky:2007hu}. Among them, exponential $f(R)$ gravity models, in which the Lagrangian includes an exponential correction to the Ricci scalar, have attracted particular attention due to their ability to interpolate smoothly between GR at high curvature (early Universe) and modified gravity at low curvature (late Universe) \cite{Cognola:2007zu, Linder:2009jz,Elizalde:2010ts,Bamba:2010ws}. In this framework, $\Lambda$CDM is recovered asymptotically at high redshifts, while dynamical deviations contribute at late times, leading to an agreement with low-redshift data. Exponential $f(R)$ gravity has been tested and constrained with a combination of various datasets \cite{Chen:2014tdy,Odintsov:2017qif}, as well as its possible extension with a generalized exponential factor \cite{Odintsov:2024woi,Odintsov:2025jfq} or the presence of axion dark matter \cite{Odintsov:2023nqe,Odintsov:2024lid}. Moreover, by including higher-order curvature terms, exponential gravity can also describe the inflationary epoch, remaining consistent with Planck constraints \cite{Odintsov:2017qif}, thereby offering a unified cosmological description from inflation to dark energy domination.

In addition, $f(R)$ gravity models can be reformulated as effective dark energy scenarios with an evolving equation-of-state parameter, providing a natural framework to test phenomenological dark energy parametrizations. Several studies have investigated the potential of modified gravity to address current cosmological discrepancies, particularly the $H_{0}$ tension \cite{dainottiApJ-H0(z),Odintsov:2020qzd,DainottiGalaxies-H0(z),Schiavone:2022wvq,Nojiri:2022ski,Montani:2023xpd,Escamilla:2024xmz,Montani:2024xys,Montani:2025jkk}. 


In parallel, massive neutrinos have become an essential ingredient in precision cosmology \cite{Lesgourgues:2014zoa,DIVALENTINO2026286}. The discovery of neutrino oscillations confirms that at least two neutrino species have non-zero masses, implying a non-negligible contribution to the total matter density of the Universe. While the standard model assumes three effectively massless neutrinos, oscillation experiments establish a lower bound on the total neutrino mass, $\sum m_\nu \geq 0.06\, \textrm{eV}$ \cite{deSalas:2020pgw,Esteban:2024eli}. A robust cosmological detection of $\sum m_\nu$ would be crucial both for particle physics and cosmology.

In the early Universe, neutrinos were thermally coupled to the primordial plasma through weak interactions and decoupled at $T\sim2 \textrm{MeV}$, forming the Cosmic Neutrino Background (C$\nu$B). Non-instantaneous decoupling and finite-temperature quantum electrodynamics (FT-QED) corrections lead to a slight excess in the effective number of relativistic species, with the standard model predicting $N_{\textrm{eff}}=3.044$ \cite{Froustey:2020mcq,Bennett:2020zkv,Drewes:2024wbw}. Massive neutrinos affect both the background expansion and the growth of cosmic structures: they behave as radiation at early times, contributing to the total radiation energy density, and later they become non-relativistic and behave as a hot dark matter component that suppresses the growth of matter fluctuations below the neutrino free-streaming scale. These effects leave measurable imprints on the CMB anisotropies, the matter power spectrum, weak lensing, and large-scale structure surveys .

Progressively tighter constraints on $\sum m_\nu$ have been obtained from cosmological data, with recent analyses combining Planck and large-scale structure datasets reaching the sub-eV level \cite{Vagnozzi:2017ovm,Planck_2020,DiValentino:2021hoh} . The latest results from DESI provide some of the most stringent cosmological bounds to date $\sum m_\nu <0.064\textrm{eV}$ (95$\%$ C.L.) \cite{DESI:2025zgx,Elbers:2025vlz} within $\Lambda$CDM, complementing terrestrial limits such as those from the KATRIN experiment, which currently reports $\sum m_\nu<0.45 \textrm{eV}$ (90$\%$ C.L.) \cite{KATRIN:2024cdt}. Interestingly, the upper limit provided by DESI changes to $\sum m_\nu <0.163\textrm{eV}$ (95$\%$ C.L.) in the $w_0w_a$CDM model \cite{DESI:2025zgx,Elbers:2025vlz}, showing that alternatives to $\Lambda$ may alleviate the neutrino mass tension between cosmological and neutrino oscillations constraints.

In this work, we investigate the cosmological implications of a generalized exponential $f(R)$ gravity model \cite{Odintsov:2024woi} in the presence of massive neutrinos, assuming a transition from relativistic to non-relativistic regimes. Our aim is to assess whether this combined framework can reproduce current cosmological observations while partially alleviating existing tensions, particularly the $H_0$ tension and the neutrino mass bounds. A single modification to $\Lambda$CDM may not be sufficient to resolve all these issues (see the discussion in \cite{Vagnozzi:2023nrq}); however, the combined effects of modified gravity and massive neutrinos could provide a viable pathway toward consistency. The interplay between $f(R)$ gravity and massive neutrinos can modify cosmological dynamics and the interpretation of observational data, hence offering new insights into the origin of cosmic acceleration and the nature of dark energy. While previous studies have investigated the interplay between massive neutrinos and specific $f(R)$ models, such as the Hu-Sawicki model \cite{Geng:2014yoa}, focusing primarily on structure formation and the matter power spectrum, the present work differs in both the theoretical framework and the scientific objectives. We consider the generalized exponential $f(R)$ model \cite{Odintsov:2024woi}, which introduces two independent parameters $\alpha$ and $\beta$ providing a more flexible description of late-time deviations from $\Lambda$CDM, and we focus on the background expansion history constrained against the latest DESI DR2 BAO data. The novelty of this analysis lies in the combined investigation of a theoretically motivated modified gravitational sector and a non-minimal matter sector through massive neutrinos, whose interplay may offer a viable pathway toward a consistent resolution of current cosmological tensions that neither modification alone can achieve.
To this end, we perform a joint analysis using multiple observational datasets, including SNe Ia, BAO, cosmic chronometers measurements, and CMB likelihoods. We focus on the late-time evolution of the Universe and compare the modified background dynamics with those of the $\Lambda$CDM model.

The paper is organized as follows. In Sec.~\ref{sec:f(R)-modified-gravity}, we review the cosmological formalism of $f(R)$ gravity. In Sec.~\ref{sec: exp_f_nu}, we introduce the generalized exponential gravity model with massive neutrinos and discuss its theoretical properties and effects on the expansion history. Sec.~\ref{sec:obs_data} presents the observational datasets employed in our analysis, while Sec.~\ref{sec:constraints} details our methodology and provides observational constraints on both the $\Lambda$CDM and modified gravity scenarios. Finally, Sec.~\ref{sec:conclusions} summarizes our main conclusions.

\section{$f(R)$ modified gravity and cosmology} \label{sec:f(R)-modified-gravity}

The gravitational Lagrangian density $\mathcal{L}_g$ can generally be constructed as a function of curvature invariants, if the resulting theory satisfies key viability conditions: general covariance, the existence of a Newtonian limit, a consistent causal structure, and light-cone compatibility. The simplest extension of General Relativity (GR) within this framework is obtained when $\mathcal{L}_g$ is a function $f(R)$ of the Ricci scalar $R$. This introduces an additional dynamical degree of freedom beyond GR \cite{Buchdahl:1970ynr}. The action for $f(R)$ gravity reads \cite{Nojiri:2006ri,Sotiriou:2008rp,Nojiri:2010wj,Capozziello:2011et,Nojiri:2017ncd}
\begin{equation}
    S=\frac{1}{2\,\kappa^2}\,\int d^{4}x\,\sqrt{-g}\,f\left(R\right)+S_{M}\,,
    \label{eq:action-f(R)}
\end{equation}
where $\kappa^2=8\pi G$ is the Einstein constant, $G$ is Newton’s gravitational constant, $g$ is the determinant of the metric tensor $g_{\mu\nu}$, and $S_M$ denotes the action for the matter fields.

Varying the action~\eqref{eq:action-f(R)} with respect to the metric yields the modified field equations in the metric formalism: 
\begin{align}
    f_{R}\left(R\right)\,R_{\mu\nu}-\frac{1}{2}\,g_{\mu\nu}\,f\left(R\right)&+g_{\mu\nu}\,\Box f_{R}\left(R\right)\nonumber\\
    &-\nabla_{\mu}\nabla_{\nu}f_{R}\left(R\right)=\kappa^2\,T_{\mu\nu}\,,
    \label{eq:field-equations-f(R)}
\end{align}
where $\nabla_{\mu}$ denotes the covariant derivative, $\Box=g^{\rho\sigma},\nabla_{\rho}\nabla_{\sigma}$ is the d'Alembert operator, $T_{\mu\nu}$ is the energy-momentum tensor of matter, and $f_{R}(R) \equiv df/dR$.

In GR, choosing a linear dependence of the Lagrangian on $R$ leads to second-order differential equations in the metric components $g_{\mu\nu}$. However, this choice is primarily motivated by simplicity rather than physical necessity. In general, the $f(R)$ field equations~\eqref{eq:field-equations-f(R)} consist of ten independent fourth-order partial differential equations in the metric components. They reduce to the Einstein field equations when $f(R) = R$.

As in GR, the conservation laws obtained from $\nabla_\nu T^{\mu\nu}=0$ follow from the Bianchi identities and remain valid in the $f(R)$ metric formalism. This can be directly verified from Eq.~\eqref{eq:field-equations-f(R)}.

For cosmological applications, it is useful to rewrite Eq.\eqref{eq:field-equations-f(R)} in an effective GR-like form:
\begin{equation}
    R_{\mu\nu}-\frac{1}{2}g_{\mu\nu}R=\kappa^2_{\text{eff}}\,\left(T_{\mu\nu}+T_{\mu\nu}^{\text{eff}}\right)\,,
    \label{eq:effective-GR-shape}
\end{equation}
hightlighting the contribution of additional geometric terms. From comparison with Eq.\eqref{eq:field-equations-f(R)}, one identifies an effective Einstein constant
\begin{equation}
    \kappa^2_{\text{eff}}\equiv \frac{\kappa^2}{f_{R}\left(R\right)}\,, \label{eq:effective-Einstein-constant}
\end{equation}
or equivalently an effective gravitational coupling $G_{\text{eff}} \equiv G / f_{R}(R)$, and an effective energy-momentum tensor:
\begin{equation}
    T_{\mu\nu}^{\text{eff}} \equiv \frac{1}{\kappa^2}\,\left[ \frac{1}{2}\,g_{\mu\nu}\,\left(f-f_{R}\,R\right)+\nabla_{\mu}\nabla_{\nu}f_{R}-g_{\mu\nu}\,\Box f_{R}\right] \,.
\end{equation}
These quantities originate from the geometrical modification introduced in the gravitational action~\eqref{eq:action-f(R)}. 

We now focus on a spatially flat (zero curvature), homogeneous, and isotropic Universe described by the Friedmann-Lema\^itre-Robertson-Walker (FLRW) metric in terms of the cosmic time and spherical coordinates $(t, r, \theta, \phi)$ \cite{Planck_2020,Efstathiou:2020wem}:
\begin{equation}
    ds^2=-dt^2+a^2(t) \left[dr^2+r^2\left(d\theta^2+\sin^2\theta d\phi^2\right)\right]\,,
    \label{eq:metric-FLRW}
\end{equation}
where $a(t)$ is the scale factor. In this background, the Ricci scalar is given by:
\begin{equation}
    R=6\left(\dot{H}+2H^2\right)
    \label{eq:expression-R}
\end{equation}
where $H(t) \equiv \dot{a}/a$ is the Hubble parameter, and the dot denotes a derivative with respect to cosmic time. We assume the Universe is filled with a perfect fluid with isotropic pressure $p$ and energy density $\rho$, and the respective stress-energy tensor is:
\begin{equation}
    T_{\mu\nu}=\left(\rho+p\right)u_\mu u_\nu+p g_{\mu\nu}\,,
\end{equation}
where $u_\mu=(1,0,0,0)$ is the four-velocity of the fluid with respect to a comoving observer. The conservation law $\nabla_\mu T^{0\mu}=0$ in the FLRW metric is
\begin{equation}
    \dot{\rho}+3H\left(\rho+p\right)=0\,,
    \label{eq:continuity}
\end{equation}
which is the standard continuity equation related to the energy conservation. In a homogeneous and isotropic Universe, all background quantities depend solely on $t$.

In this context, the modified gravitational field equations~\eqref{eq:field-equations-f(R)} reduce to
\begin{align}
    & H^{2} f_{R}=\frac{\kappa^2\rho}{3}+\frac{R\,f_{R}-f}{6}-\dot{f}_{R}\,H \label{eq:modified-Friedmann}\,,\\
    & f_{R} \left(\frac{R}{6}+H^2\right)-\frac{f}{2}-\ddot{f}_R-2H\dot{f}_R=\kappa^2 p \label{eq:modified-11}\,,
\end{align}
corresponding to the $tt$ and $rr$ components, respectively\footnote{The $rr$ component of the modified gravitational field equations can appear in several equivalent forms. Indeed, using Eq.\eqref{eq:expression-R} and the definition of $H$, the term $\frac{R}{6}+H^2$ in Eq.~\eqref{eq:modified-11} can be rewritten equivalently as $\dot{H}+3H^2$, $\ddot{a}/a+2H^2$, $\frac{R}{3}-\frac{\ddot{a}}{a}$, or $\frac{R}{2}-2\dot{H}-3H^2$. Moreover, by applying the chain rule, the derivatives of $f_R$ can be expressed as $\dot{f}_{R}=f_{RR} \dot{R}$ and $\ddot{f}_R=f_{RRR} \left(\dot{R}\right)^2+f_{RR}\ddot{R}$.}. These equations are not independent: any one can be derived from the other one by using the continuity equation~\eqref{eq:continuity}, as in GR.

By combining Eqs.~\eqref{eq:modified-Friedmann} and \eqref{eq:modified-11}, the modified acceleration equation is obtained:
\begin{equation}
    \frac{\ddot{a}}{a} f_R = -\frac{\kappa^2}{6}\left(\rho+3p\right)+\frac{R\,f_{R}-f}{6}-\frac{1}{2}\left(H \dot{f}_R+\ddot{f}_R\right)\,.
    \label{eq:modified-acceleration}
\end{equation}
To better compare the cosmic dynamics in $f(R)$ gravity with the standard $\Lambda$CDM model, we recast Eqs.~\eqref{eq:modified-Friedmann} and \eqref{eq:modified-acceleration} in the effective form
\begin{align}
    & H^2 = \frac{\kappa^2_{\text{eff}}}{3} \left(\rho + \rho_{\text{eff}}\right) \label{eq:Friedmann-effective}\,,\\
    & \frac{\ddot{a}}{a}=-\frac{\kappa^2_{\text{eff}}}{6}\left(\rho + \rho_{\text{eff}}+3p+3p_{\text{eff}}\right)\,,\label{eq:acceleration-effective}
\end{align}
according to the idea presented in Eq.~\eqref{eq:effective-GR-shape}. We have introduced the effective pressure and energy density:
\begin{align}
    \rho_{\text{eff}} &= \frac{R f_R - f}{2 \kappa^2} - \frac{3 H \dot{f}_R}{\kappa^2} \label{eq:rho_eff}\,,\\
    p_{\text{eff}} &= \frac{\ddot{f}_R}{\kappa^2}-\frac{R f_R - f}{2 \kappa^2} + \frac{2 H \dot{f}_R}{\kappa^2} \,. \label{eq:p_eff}
\end{align}
These effective quantities act as a geometrically induced fluid, providing a possible alternative to dark energy. In this framework, cosmic acceleration ($\Ddot{a}>0$) can be interpreted as arising from curvature effects rather than a true cosmological constant.

For convenience, we now switch to the redshift variable $z$, defined via $1+z = a_0/a(t)$, with $a_0 = 1$ by convention. From this, and using $H(t) = \dot{a}/a$, it follows that $dz/dt = - (1+z) H(z)$. We thus express Eqs.~\eqref{eq:expression-R}, \eqref{eq:modified-Friedmann}, and \eqref{eq:continuity} in terms of $z$:

\begin{align}
    & H^\prime(z)=\frac{1}{1+z} \left(2H-\frac{R}{6H}\right)\label{eq:expression-R-in-z}\,,\\
    & R^\prime(z)= - \frac{1}{\left(1+z\right) f_{RR}} \left(\frac{\kappa^2 \rho}{3 H^2} - f_R + \frac{R f_R -f}{6 H^2}\right)\,,\label{eq:Friedmann-in-z}\\
    & \left(1+z\right) \rho^\prime (z) - 3 \left(\rho+p\right)=0\,,\label{eq:continuity-in-z}
\end{align}
respectively, where primes denote derivatives with respect to $z$. We also used the chain rule $f_R^\prime = f_{RR} R^\prime$, assuming $f = f(R)$ and $R = R(z)$.


\section{Generalized exponential gravity in the presence of neutrinos}\label{sec: exp_f_nu}

Thus far, we have not specified the functional form of $f(R)$, which plays a central role in determining the cosmological dynamics and observables. In what follows, we focus on a particular class known as exponential $f(R)$ gravity \cite{Cognola:2007zu,Linder:2009jz,Elizalde:2010ts,Bamba:2010ws,Odintsov:2017qif}. Specifically, we consider the extended $f(R)$ model presented in \cite{Odintsov:2024woi}. Additionally, here we include neutrino contribution with a transition from relativistic to non-relativistic particles to account for the consequences on cosmological dynamics.

The gravitational Lagrangian density of the generalized exponential $f(R)$ gravity is defined as:
\begin{equation}
    f(R)=R - 2\Lambda \left\{1-\exp{\left[-\beta \left(\frac{R}{2\Lambda}\right)^\alpha\right]}\right\}\,,
    \label{eq:exponential-f(R)-gravity}
\end{equation}
where $\alpha$ and $\beta$ are dimensionless parameters, and $\Lambda$ is a constant. Introducing the dimensionless Ricci scalar $\mathcal{R} \equiv R/(2\Lambda)$, the function can be rewritten as:
\begin{equation}
    f(\mathcal{R})=2\Lambda \left(\mathcal{R} -1+e^{-\beta \mathcal{R}^\alpha}\right)\,.
    \label{eq:exponential-f(R)-gravity-dimensionless}
\end{equation}
This model generalizes standard exponential gravity \cite{Odintsov:2017qif}, which corresponds to the case $\alpha=1$. In the limit $\beta \rightarrow +\infty$ and/or $\mathcal{R} \rightarrow +\infty$ with $\alpha>0$, the exponential term becomes negligible, and $f(R)$ reduces to the Einstein-Hilbert form with a cosmological constant $\Lambda$, effectively recovering the $\Lambda$CDM model.

The dimensionless parameters $\alpha$ and $\beta$ govern the shape and strength of the exponential modification to the gravitational Lagrangian. Specifically, $\beta$ controls the overall amplitude of the deviation from GR: larger values of $\beta$ suppress the exponential correction more rapidly, driving the model toward the $\Lambda$CDM limit, while smaller values allow for more pronounced modifications at low curvature. The parameter $\alpha$ determines the curvature scale at which the transition from the GR regime to the modified gravity phase occurs: for $\alpha > 1$, the exponential term is suppressed at high curvature more efficiently, ensuring a faster recovery of the $\Lambda$CDM behavior at early times, whereas $\alpha < 1$ broadens the transition and extends the modified gravity effects to higher redshifts. Together, $\alpha$ and $\beta$ effectively parameterize the onset and the depth of the dark energy-like phase induced by the geometric modification, making them the key parameters controlling the late-time deviations from standard cosmology.

Substituting Eq.~\eqref{eq:exponential-f(R)-gravity-dimensionless} into Eqs.~\eqref{eq:Friedmann-effective}, \eqref{eq:acceleration-effective}, \eqref{eq:rho_eff}, and~\eqref{eq:p_eff}, one finds that this model can mimic a dark energy component at late times. Furthermore, exponential gravity can be extended by including additional terms to account for the early Universe inflationary phase, dominated by large curvature ($\mathcal{R} \gg 1$), while remaining consistent with observational constraints from Planck data \cite{Planck_2020}. In other words, exponential gravity may unify the cosmological description from the inflation to the dark energy dominated Universe \cite{Odintsov:2017qif}. In this work, however, we restrict our attention to the post-recombination era, where inflationary corrections are negligible.

\begin{figure}[htb]
    \centering
    \includegraphics[width=1\linewidth]{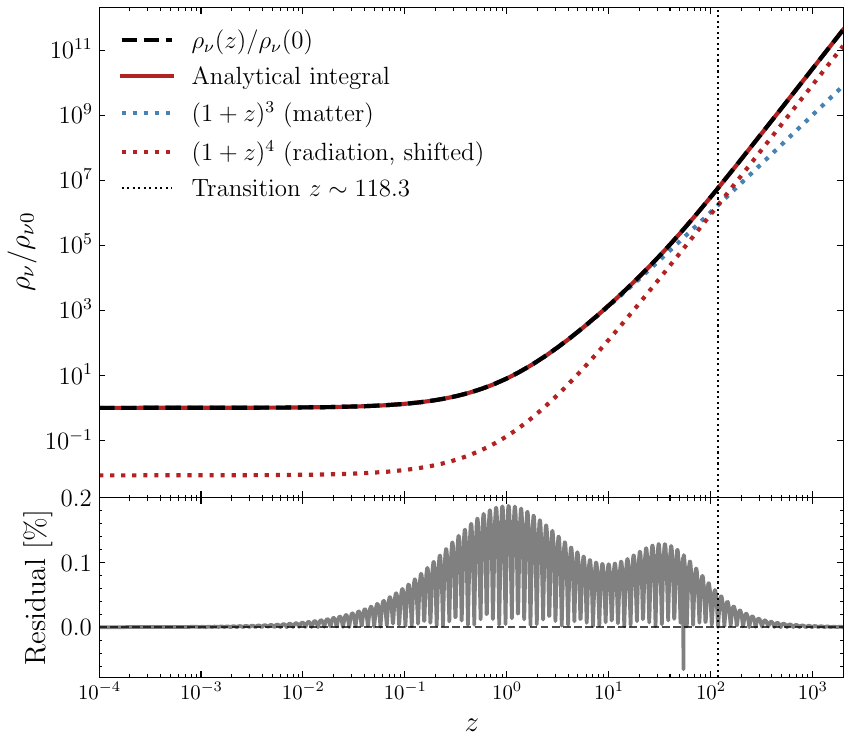}
    \caption{Numerical evaluation of the Fermi-Dirac integral \eqref{eq:fermi-dirac} for fixed $\sum m_\nu = 0.06$ eV. The vertical line represents the transition redshift from relativistic to non-relativistic neutrinos. The energy density behaves as the one for a non-relativistic (blue) specie at low redshift and as a relativistic (red) one at high redshifts. We also report the analytical integral of the Fermi-Dirac distribution. In the lower panel, we show the residuals in percentage with respect to the interpolation function, demonstrating good agreement across all epochs.}
    \label{fig:rho_nu}
\end{figure}

We assume the cosmological fluid consists of the following components: pressure-less matter ($p_m = 0$), including both cold dark matter ($\rho_c$) and baryons ($\rho_b$); a radiation component (photons and relativistic particles except for neutrinos) with $p_r = \rho_r / 3$; the contribution arising from all neutrino species\footnote{We explicitly separate the neutrino contribution from the matter and radiation components to investigate their role in different cosmic eras. Therefore, the matter and radiation components do not include neutrinos.}. Since the conservation laws remain valid in $f(R)$ gravity, the redshift evolution of the energy density $\rho = \rho_\text{m} + \rho_\text{r} + \rho_\nu$ follows
\begin{equation}
    \rho(z) = \rho_{\text{m}0}\,(1+z)^3 + \rho_{\text{r}0}\,(1+z)^4 + \rho_\nu(z) \ ,
\end{equation}
as obtained from the continuity equation~\eqref{eq:continuity-in-z} applied to each component separately. We have assumed that there are no interactions between different components. Hereinafter, we use the subscript 0 to denote the present-day value of a cosmological quantity. The massive neutrino energy density $\rho_\nu(z)$ can be computed from the phase-space integral over the Fermi–Dirac distribution \cite{Elbers:2025vlz}:
\begin{equation}\label{eq:fermi-dirac}
    \rho_\nu(z) = \sum_i^{N_\nu}\frac{g_i}{2\pi^2}(1+z)^4\int_0^\infty dq\frac{q^2\sqrt{q^2+a^2 m_i^2}}{1+e^{q/T_{\nu0}}} \ ,
\end{equation}
where $T_{\nu0}$ is the present-day temperature of neutrinos and $g_i$ is the degeneracy factor. The square root contains the relativistic expression for the particle energy in an expanding Universe. Considering three massive neutrino species, the effective mass is simply defined as $m = \sum m_i/3$, and $T_{\nu0} = (4/11)^{1/3} \, T_\text{CMB}$.
We set $g_i=2$ for each active neutrino mass eigenstate, corresponding to the two thermally populated helicity/chirality states (neutrino and antineutrino). This choice is valid both in the relativistic and non-relativistic regimes, unless additional states (e.g. thermally populated right-handed sterile states) are introduced (not considered in this scenario).
Actually, assuming that all neutrinos belong to the same species, the ratio $\rho_\nu(z)/\rho_\nu(0)$, and then the analysis, is independent of the values of the individual $g_i$. As an example, we show the energy density of neutrinos $\rho_\nu(z)/\rho_\nu(0)$ in Fig. \ref{fig:rho_nu} for $\sum m_\nu=0.06$ eV.

In general, the cosmological parameters in exponential $f(R)$ gravity, such as the Hubble constant $H_0$ and the present-day density parameters, deviate from their counterparts in the flat $\Lambda$CDM model, denoted by a superscript $*$. So that
\begin{equation}
         H_0\neq H_0^*\, \qquad \text{and} \qquad \Omega_i\neq \Omega_i^* \ ,
\end{equation}
where\footnote{We drop the subscript $0$ in the cosmological parameters $\Omega_i$ for notational convenience.}
\begin{equation}
    \Omega_i \equiv \frac{\rho_{i0}}{\frac{3}{\kappa^2}{H_0^2}} \quad\text{and}\quad\Omega_i^* \equiv \frac{\rho_{i0}}{\frac{3}{\kappa^2}{H^*_0{}^2}}  
\end{equation}
for $i =\text{m},\text{r}$ and $\Lambda$,
while the fractional present-day energy densities of neutrinos are 
\begin{equation}
    \Omeganu \equiv \frac{1}{h^2} \frac{\sum m_\nu}{93.14 \text{ eV}} \quad\text{and}\quad \Omeganu^* \equiv \frac{1}{h^{*2}} \frac{\sum m_\nu}{93.14 \text{ eV}}\,.
\end{equation}
Here, $h$ are the usual dimensionless Hubble constant defined as $h \equiv H_0[\textrm{km s}^{-1} \textrm{Mpc}^{-1}]/100\,$.
In the flat $\Lambda$CDM model the number of parameters can be always reduced by evaluating the standard Friedmann equation for $z=0$, providing the well-known relation:
\begin{equation}\label{eq:vincolo_LCDM}
    \Omega^*_\Lambda = 1- \Omegam^*-\Omegar^* - \Omeganu^* \ .
\end{equation}
Conversely, in $f(R)$ gravity, the modified Friedmann equation \eqref{eq:Friedmann-in-z} for $z=0$ can be written as
\begin{equation}
    \Omega_\Lambda = \Omega_{f(R)}^0 - \Omegam-\Omegar-\Omeganu \,,
\end{equation} 
where
\begin{equation}
    \Omega_{f(R)}^0 = \bigg(f_R- \frac{R f_R -f}{6 H^2}  -R^\prime f_{RR}\bigg)\Bigg|_{z=0} +\Omega_\Lambda 
\end{equation}
represents the extra fractional energy density arising from the modified gravity and can always be shifted by a constant term. This contribution $\Omega_{f(R)}^0$ cannot be computed analytically due to the non-minimal coupling between the function $f_R$ and the metric. In other words, we cannot write a simple relation between the parameters $\Omega_i$, because we would need an expression of the Ricci scalar and the function $f_R$ independent of $\Omega_i$, whereas the relation \eqref{eq:expression-R} for the Ricci scalar and the modified Friedmann equations show this non-minimal coupling.

\begin{figure}[htb!]
\begin{center}
\centering
\includegraphics[width=1\linewidth]{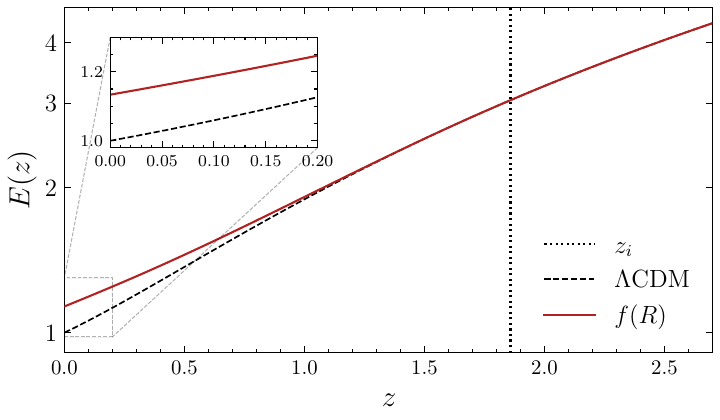}
\caption{Plot of the function $E(z)$ defined in Eq.~\eqref{eq:expression-E(z)} (red curve) compared with the respective dimensionless Hubble function in the $\Lambda$CDM model (dashed black curve). The vertical black dotted line indicates the initial redshift $z_i$, defined in Eq.~\eqref{eq:initial-redshift}, below which ($z<z_i$) the deviations between the exponential $f(R)$ cosmology and the $\Lambda$CDM model become non-negligible. In this plot we used the best fit parameters of the full analysis given in Table \ref{table:Best_Fit_Table}.}
\label{fig:E_z_F(r)}
\end{center}
\end{figure}

Therefore, to simplify the following numerical computations and considering that we want to recover the $\Lambda$CDM limit for high redshifts, we express the cosmological parameters of the $f(R)$ model in terms of the respective $\Lambda$CDM parameters. To this end, we define the dimensionless Hubble parameter $E(z)$ as in \cite{Odintsov:2024woi}
\begin{equation}
    E(z) \equiv \frac{H(z)}{H_0^*} \label{eq:expression-E(z)}\ .
\end{equation}
The function $E(z)$ has been adimensionalized through the Hubble constant $H_0^*$ defined within the $\Lambda$CDM framework, in this way the fractional energy densities $\Omega_i^*$ satisfy the constraint \eqref{eq:vincolo_LCDM}, making the system well defined. Then, for each choice of the parameters, we can pass from one set of variables to the other just computing the value of $E(0)$. Indeed, from Eq.~\eqref{eq:expression-E(z)}, it is trivial to write the relation between the Hubble constants: $H_0 = H_0^*E(0)$. Then, from this relation, all the other cosmological parameters are easily obtained as
\begin{equation}
    \Omega_i \equiv \frac{\rho_{i0}}{\frac{3}{\kappa^2}{H_0^2}} = \frac{1}{E(0)^{2}}\frac{\rho_{i0}}{\frac{3}{\kappa^2}{H^*_0{}^2}} = \frac{\Omega_i^*}{E(0)^2} \ . \label{eq:mapping-relations}
\end{equation}
Note that $E(0) \geq 1$ since there is a mismatch between the function $H(z)$ in $f(R)$ gravity and the Hubble constant $H_0^*$ in $\Lambda$CDM (the equality $E(0) = 1$ holds only if we consider the cosmological evolution in the $\Lambda$CDM model). This fact can be seen in Fig. \ref{fig:E_z_F(r)}. As a consequence, $H_0 \geq H_0^*$ and we also see from Eq.~\eqref{eq:mapping-relations} that $\Omega_i \leq \Omega_i^*$.
However, for high redshifts, we would expect to recover the $\Lambda$CDM limit.\\
Now we focus on the modified cosmological dynamics in $f(R)$ gravity by adopting the respective $\Lambda$CDM parameters. For numerical purposes, we write the redshift-space equations ~\eqref{eq:expression-R-in-z} and~\eqref{eq:Friedmann-in-z} in terms of dimensionless quantities $E(z)$ and $\mathcal{R}(z)$:
\begin{widetext}
\begin{align}
    E^\prime &= \frac{1} {1+z} \left(2E - \Omega_{\Lambda}^* \frac{\mathcal{R}}{E}\right)\label{eq:final1}\\
    \mathcal{R}^\prime &= \frac{E^2 \left(1-\alpha \beta \mathcal{R}^{\alpha-1}e^{-\beta \mathcal{R}^\alpha}\right)-\Omegam^* (1+z)^3 - \Omegar^* (1+z)^4 - \Omeganu^* \frac{\rho_{\nu}(z)}{\rho_{\nu}(0)} -\Omega_{\Lambda}^* \left[1-e^{-\beta \mathcal{R}^\alpha}\left(1+\alpha \beta \mathcal{R}^\alpha\right)\right]} {\left(1+z\right)\, \alpha \,\beta \, \mathcal{R}^{\alpha-2} \, e^{-\beta \mathcal{R}^\alpha} \, \left(1-\alpha+\alpha\, \beta\, \mathcal{R}^\alpha\right)E^2}  \ .\label{eq:final2}
\end{align}
\end{widetext}
The system of differential equations generalizes Eqs. (12) and (13) of \cite{Odintsov:2024woi} in the presence of neutrinos.
The system of these two coupled differential equations can be solved numerically once we set the initial conditions. Since the proposed $f(R)$ model reproduces the $\Lambda$CDM cosmology at high redshift due to the decreasing exponential term in the $f(R)$ functional form, it is not necessary to integrate the system of differential equations over the entire cosmic history, which would otherwise be a numerically demanding and time-consuming task. Therefore, the initial conditions are given by the $\Lambda$CDM cosmology for a sufficiently high initial redshift $z_i$. Then, we can integrate the system backward to $z=0$, whereas deviations from $\Lambda$CDM are not negligible. \\
At this redshift $z_i$, we write the dimensionless Friedmann equation and the expression of the dimensionless Ricci scalar within $\Lambda$CDM for $z=z_i$:
\begin{widetext}
\begin{align}
    E^2(z_i) &= \left[\Omegam^* \left(1+z_i\right)^3 + \Omegar^* \left(1+z_i\right)^4 + \Omeganu^* \frac{\rho_{\nu}(z_i)}{\rho_{\nu}(0)} + \Omega_{\Lambda}^*\right]\label{eq:initial-condition1}\\
    \mathcal{R}(z_i) &= 2 + \frac{1}{2\Omega_{\Lambda}^*} \left[\Omegam^* \left(1+z_i\right)^3 + 4\Omega_\nu^\ast \frac{\rho_\nu(z_i)}{\rho_\nu(0)} -\left(1+z_i\right) \Omeganu^* \frac{\rho'_{\nu}(z_i)}{\rho_{\nu}(0)}\right]\,.\label{eq:initial-condition2} 
\end{align}
\end{widetext}
We used the continuity equation in redshift space \eqref{eq:continuity-in-z} to write $p_\nu(z)=\frac{1+z}{3} \rho_\nu^\prime(z) - \rho_\nu(z)$.
The initial redshift $z_i$ can be determined if we fix the deviation $\epsilon\ll1$ from $\Lambda$CDM given by the exponential factor in the gravitational Lagrangian density \eqref{eq:exponential-f(R)-gravity-dimensionless}:
\begin{equation}
    \epsilon \equiv e^{-\beta \mathcal{R}^\alpha (z_i)}\,. \label{eq:epsilon}
\end{equation}

Combining Eqs.~\eqref{eq:initial-condition2} and~\eqref{eq:epsilon}, we obtain
\begin{align}
    & 2 \Omega_\Lambda^* \left[\left(\frac{\ln\epsilon^{-1}}{\beta}\right)^{1/\alpha}-2\right] = \nonumber\\
    &\left[\Omegam^* \left(1+z_i\right)^3 + 4\Omega_\nu^\ast \frac{\rho_\nu(z_i)}{\rho_\nu(0)} -\left(1+z_i\right) \Omeganu^* \frac{\rho'_{\nu}(z_i)}{\rho_{\nu}(0)}\right] \label{eq:z_i_intermediate}
\end{align}
We assume that neutrinos are non-relativistic at the initial redshift $z_i$, implying $\rho_\nu\propto \left(1+z_i\right)^3$. This assumption is justified because deviations from $\Lambda$CDM in the exponential $f(R)$ model typically occur at redshifts lower than those at which neutrinos become non-relativistic. Consequently, it is straightforward to solve Eq.~\eqref{eq:z_i_intermediate} for $z_i$
\begin{equation}
    z_i = \left\{ \frac{2 \Omega_\Lambda^*}{\Omegam^*+\Omega^*_{\nu}} \left[\left(\frac{\log{\epsilon^{-1}}}{\beta}\right)^{1/\alpha}-2\right] \right\}^{1/3} -1 \,. \label{eq:initial-redshift}
\end{equation}

To summarize, the cosmological evolution is undistinguishable from the $\Lambda$CDM dynamics for $z\geq z_i$, whereas we need to solve numerically the system of differential equations \eqref{eq:final1} and \eqref{eq:final2} for $0\leq z\leq z_i$ to find the solution in $f(R)$ modified cosmology.  

\section{Observational datasets}\label{sec:obs_data}
In this section, we describe the various observational datasets we used to constrain cosmological parameters and their joint analysis. Specifically, we consider the following datasets: Cosmic Chronometers (CC), Baryon Acoustic Oscillations (BAO), Type Ia Supernovae (SNe Ia), and Cosmic Microwave Background (CMB). 

\begin{itemize}
    \item \textbf{Cosmic Chronometers (CC)}:\\
    We consider 31 data points in the redshift range $ 0.07 \leq z \leq 1.965 $ (see Table \ref{table:CC_Data}). The corresponding $\chi^2$ value is given by
    \begin{equation}\label{chi_CC}
        \chi^2_{\text{CC}} = \sum_{i=1}^{31}\frac{(H^\text{obs}(z_i)-H^\text{th}(z_i))^2}{\sigma^2_H(z_i)} \ ,
    \end{equation}
    where $H^\text{obs}(z_i)$ is the observed value with uncertainty $\sigma_H(z_i)$, and $H^\text{th}(z_i)$ represents the theoretical prediction given by the cosmological model.
    This dataset does not have the constraining power of the others (such as the CMB) on the standard cosmological parameters, but it will help constraining the $f(R)$ ones since it lies at at low redshift, where the effects of the dark energy are present.  

    \item \textbf{Type Ia Supernovae (SNe Ia)}: \\
    We consider the \textit{Pantheon$+$} dataset\footnote{The repository of the Pantheon$+$ sample can be found at the following link: \url{https://github.com/PantheonPlusSH0ES/DataRelease} .} \cite{Brout_2022}, labeled hereinafter as Pan+, which consist of 1550 data points in the redshift range $0.001\leq z\leq 2.26$.
    Each data point is given by the observed distance modulus at a given redshift $\mu^\text{obs}(z)$. The $\chi^2$ value can be computed as
    \begin{equation}
        \chi^2_{\text{SNeIa}} = \sum_{i,j=1}^{1550}\Delta\mu_i\Big(\text{Cov}^{-1}\Big)_{ij}\Delta\mu_j \ ,
    \end{equation}
 where $\Delta\mu_i = (\mu^\text{obs}(z)-\mu^\text{th}(z))_i$ and $\text{Cov}$ is the covariant matrix of the dataset obtained summing the statistical and systematic covariance matrices. The theoretical expression of the distance modulus is 
    \begin{equation}
    \mu^\text{th}(z) = 5 \, \log_{10}\left(\frac{D_L(z)}{10 \text{pc}}\right) \ ,
    \end{equation}
    which is related to the luminosity distance $D_L(z)$:
    \begin{equation}
        D_\mathrm{L}(z) = (1+z) \int_0^z\frac{dz'}{H(z')} \ .
    \end{equation}
    This latter quantity depends on the considered cosmological model and encodes information about the cosmological dynamics through the presence of the Hubble parameter.

    \item \textbf{Baryon Acoustic Oscillations (BAO):} \\
    We use of DESI DR2 measurements of BAO \cite{DESI:2025zgx} obtained from galaxy and quasar data. Specifically, we considered the measurements of $D_M(z) / r_d$ and $D_H(z)/r_d$ for six redshift bins from $0.1\leq z\leq4.2$, where $D_H(z) \equiv 1/H(z)$ is the Hubble distance in natural units ($c=1$), $D_M(z)$ is the transverse comoving distance, related to the luminosity distance through the relation $D_L(z)=(1+z)D_M(z)$, and $r_d$ is the sound horizon at the baryon drag epoch. The value of $r_d$ can be derived from its definition through an integral over the redshift, but following the approach of the DESI collaboration \cite{DESI:2025zgx}, we use the fitting formula \cite{Brieden_2023}
    \begin{align}  
        r_d ~=~ & 147.05\,{\rm Mpc} \,\nonumber\times\\
        &  \left(\frac{\omegab}{0.02236}\right)^{-0.13} 
               \left(\frac{\omegabc}{0.1432}\right)^{-0.23}    
            \left(\frac{\Neff}{3.04}\right)^{-0.1}\ ,
        \label{eq:rdformula}
    \end{align}
    where the standard notation is used, $\omegab = \Omegab h^2$ and $\omegabc = (\Omegab+\Omegac) h^2$.
    We consider the number of effective neutrino species to be constant $\Neff=3.04$, since varying this parameter does not produce relevant changes in dark energy analysis \cite{DESI:2025zgx}.

    \item \textbf{Cosmic Microwave Background (CMB):}\\
    We use the compression of the full information of the Planck 2018 data \cite{Planck_2020} on the parameters $(\theta_\ast,\omegab,\omega_{cb})$, based on the \texttt{CamSpec} CMB likelihood as computed in \cite{Lemos:2023xhs}. We assume $T_\text{CMB} = 2.7255$ K \cite{Fixsen:2009ug}. We use this analysis since these quantities are related to the early epoch of the Universe, in which the $f(R)$ model replicates the $\Lambda$CDM case. The value of the acoustic angular scale is defined as $\theta_\ast = r_\ast / D_M(z_\ast)$, where $z_\ast$ is the redshift at the end of recombination and $r_\ast$ is the sound horizon at this redshift. \\
    The dataset is given by the mean vector
    \begin{equation}
        \mu^\text{obs} = \begin{pmatrix}
        \theta_\ast\\
        \omegab\\
        \omegabc
        \end{pmatrix} = 
        \begin{pmatrix}
        0.01041 \\
        0.02223 \\
        0.14208
        \end{pmatrix}
        \label{eq:CMB_compression_mean}
    \end{equation}
    and covariance
    \begin{equation}
        \text{Cov} = 10^{-9} \times 
        \begin{pmatrix}
        0.006621 &  0.12444 & -1.1929 \\
        0.12444 &  21.344 & -94.001 \\
        -1.1929 & -94.001 & 1488.4
        \end{pmatrix}\,.
        \label{eq:CMB_compression_cov}
    \end{equation}
    To perform a fully rigorous analysis of the CMB dataset, one should derive the parameters $z_\ast$ and $r_\ast$ through a prior analysis. However, since we assume that the exponential $f(R)$ gravity reproduces exactly $\Lambda$CDM at high redshift, we 
    keep fixed $z_\ast = 1089$, and we used the formula\footnote{This formula has been computed for the case of fixed neutrino mass, but it can be shown that a variation of the latter does not produce relevant changes in the constant of proportionality.} for $r_\ast = r_d/1.01841$ from \cite{Hu_1996}. \\
    We stress that this compressed CMB likelihood is based on the acoustic scale $\theta_*$ and the physical baryon and cold dark matter densities, which are primarily sensitive to the background expansion history. As a consequence, this likelihood does not probe the CMB power spectra directly, and therefore does not require a perturbative treatment of the neutrino sector. Similarly, the BAO measurements constrain the comoving sound horizon and angular diameter distances, which are determined by the background evolution. The effect of massive neutrinos on the growth of structure and the CMB anisotropies, while important for full-shape analyses, is therefore not directly probed by our dataset combination.
\end{itemize}

\section{Constraints on cosmological parameters} \label{sec:constraints}
We perform a standard Bayesian parameter inference \cite{padilla2021cosmological} using a Markov Chain Monte Carlo (MCMC) approach to obtain the posterior distributions of the cosmological parameters. The analysis is carried out with the publicly available \texttt{emcee} Python package \cite{Foreman-Mackey2019}.

As discussed in Sec.~\ref{sec: exp_f_nu}, the cosmological parameters are mapped into the equivalent $\Lambda$CDM quantities through the relations in Eq.~\eqref{eq:mapping-relations}, ensuring also that the model recovers the $\Lambda$CDM limit at high redshifts. Including the additional exponential $f(R)$ parameters $\alpha$ and $\beta$, the parameter space explored in the MCMC analysis is $(H^\ast_0, \alpha,\beta,\omegab^\ast,\omegabc^\ast,\omega_\nu^\ast)$. For each MCMC sample, we numerically solve the system of differential equations \eqref{eq:final1} and \eqref{eq:final2}, which is notably stiff due to the exponential terms in the $\mathcal{R}'$ equation. The initial redshift $z_i$, defined in Eq.~\eqref{eq:initial-redshift}, is chosen to guarantee a slight deviation from the $\Lambda$CDM solution by fixing $\epsilon=10^{-7}$. The corresponding initial conditions are set by Eqs.~\eqref{eq:initial-condition1} and \eqref{eq:initial-condition2}. We integrate the equations backward from $z_i$ to $z=0$ using the Radau integration method and match the resulting solution to the $\Lambda$CDM one in the range $z_i<z<z_\ast$. This approach significantly reduces numerical instabilities and computational cost.

The following uniform priors are adopted for the sampled parameters:
\begin{align}
    H^\ast_0 & \in [40,100]\nonumber\\
    \alpha& \in [0.01,5]\nonumber\\
    \beta& \in [0.01,5]\nonumber\\
    \omegab^\ast& \in [0.001,1]\nonumber\\
    \omegabc^\ast& \in [0.001,1]\nonumber\\\omega_\nu^\ast& \in [0,1]\nonumber \ ,
\end{align}
where the Hubble parameter is given in $[\text{km}\,\text{s}^{-1}\,\text{Mpc}^{-1}]$.
As discussed, the model asymptotically recovers $\Lambda$CDM in the limit $\beta\rightarrow\infty$. Setting the upper prior boundary at $\beta = 5$
 yields deviations from $\Lambda
$CDM of only $\mathcal{O}(10^{-8}\%)$, thereby ensuring a sufficiently wide prior range for this parameter.
To reduce the number of parameters of the sampling, the present radiation density is fixed to $\Omegar =  2.469\times10^{-5} h^{-2}$, and we set the effective number of neutrino species to be $\Neff = 3.044$. As shown in \cite{DESI:2025zgx}, this latter assumption has negligible impact on the results. 
After obtaining the posterior distributions for $(H^\ast_0, \alpha,\beta,\omegab^\ast,\omegabc^\ast,\omega_\nu^\ast)$, we recover the original physical parameters of the $f(R)$ model $(H_0, \alpha,\beta,\omegab,\omegabc,\omega_\nu)$ using again the mapping in Eq.~\eqref{eq:mapping-relations}.

We analyze two combinations of datasets, as described in Sec.~\ref{sec:obs_data}: 1) CC, DESI DR2, and CMB-$\theta_\ast$; 2) the same combination with the addition of the Pan+ SNe Ia catalog. The inclusion of late-time probes such as SNe Ia is crucial for constraining $f(R)$ models, since deviations from $\Lambda$CDM become significant at low redshifts. Meanwhile, CMB data help ensure the correct high-redshift behavior and improve convergence of the MCMC chains. For comparison, we also analyze the flat $\Lambda$CDM model and the CPL parametrization \cite{Chevallier:2000qy,Linder:2002et} for dark energy, or $w_0w_a$CDM model, where $w(z)=w_0+w_a\frac{z}{1+z}$ with $w_0$ and $w_a$ constant.

\setlength{\arrayrulewidth}{0.1mm}
\setlength{\tabcolsep}{4pt}
\renewcommand{\arraystretch}{1.5}
\begin{table*}[htb]
\centering
\begin{tabular}{|c || c | c | c | c | c | c | c | c |} 
\hline
  \multicolumn{9}{|c|}{CC \& DESI \& CMB }\\
  \hline
   Model & $\min\chi^2/\text{dof}$ & $\Delta$AIC  & $\Delta$BIC & $H_0\,\, [\text{km}\,\text{s}^{-1}\,\text{Mpc}^{-1}]$ & $\Omegam$ & $\sum m_\nu$ [eV]& $\alpha$ & $\beta$ \\
  \hline
    $\Lambda$CDM & $0.591$ & $0$ & $0$ & $68.55 \pm 0.36$ & $0.2986 \pm 0.0040$ & $<0.0536$ &  &  \\
    $w_0w_a$CDM & $0.544$  & $-1.96$ & $-5.39$ & $64.71 \pm 2.37$ & $0.3397 \pm 0.0256$ & $<0.140$  &  &  \\
    $e^{-\beta\mathcal{R}^\alpha}$ & $0.702$ & $-8.63$ & $-12.06$ & $67.30 \pm 1.28$ & $0.3112 \pm 0.0126$ & $<0.0590$ & $1.31^{+0.31}_{-0.79}$ & $1.28^{+0.40}_{-1.10}$ \\
  \hline
  \hline
  \multicolumn{9}{|c|}{CC \& DESI \& CMB \& Pan+}\\
  \hline
  Model & $\min\chi^2/\text{dof}$  & $\Delta$AIC& $\Delta$BIC & $H_0\,\, [\text{km}\,\text{s}^{-1}\,\text{Mpc}^{-1}]$ & $\Omegam$ & $\sum m_\nu$  [eV]& $\alpha$ & $\beta$ \\
  \hline
    $\Lambda$CDM & $1.067$ & $0$ & $0$ & $71.43 \pm 0.13$ & $0.2693 \pm 0.0016
$ & $<0.0126$    &  &  \\
    $w_0w_a$CDM & $1.041$ & $43.75$ & $32.84 $ & $71.13 \pm 1.20$ & $0.2810 \pm 0.104$ & $<0.0613$   &  &  \\
    $e^{-\beta\mathcal{R}^\alpha}$ & $1.054$  & $20.46$ & $9.55$ & $70.89\pm0.19$ & $0.2810^{+0.0021}_{-0.0024}$ & $<0.0290$ & $2.00\pm0.13$ & $0.294^{+0.082}_{-0.130}$ \\
  \hline
\end{tabular} 
\caption{Best-fit values of the cosmological parameters for the $\Lambda$CDM, $w_0w_a$CDM, and exponential $f(R)$ models. The upper panel reports results from the joint analysis of CC, DESI DR2, and CMB-$\theta_\ast$ data, while the bottom panel includes the Pan+ SNe Ia sample. The first column lists the considered cosmological model; the next three columns display, respectively, the reduced $\chi^2$, the relative difference $\Delta$AIC and $\Delta$BIC with respect to $\Lambda$CDM, used to assess the statistical preference for alternative cosmologies. The next three columns show the best-fit values of the parameters common to all models. The following two columns present the additional parameters specific to the exponential $f(R)$ model. All uncertainties correspond to $1\sigma$ ($\sim68\%$ C.L.).}
\label{table:Best_Fit_Table}
\end{table*}

\begin{figure}[htb!]
\begin{center}
\centering
\includegraphics[width=1\linewidth]{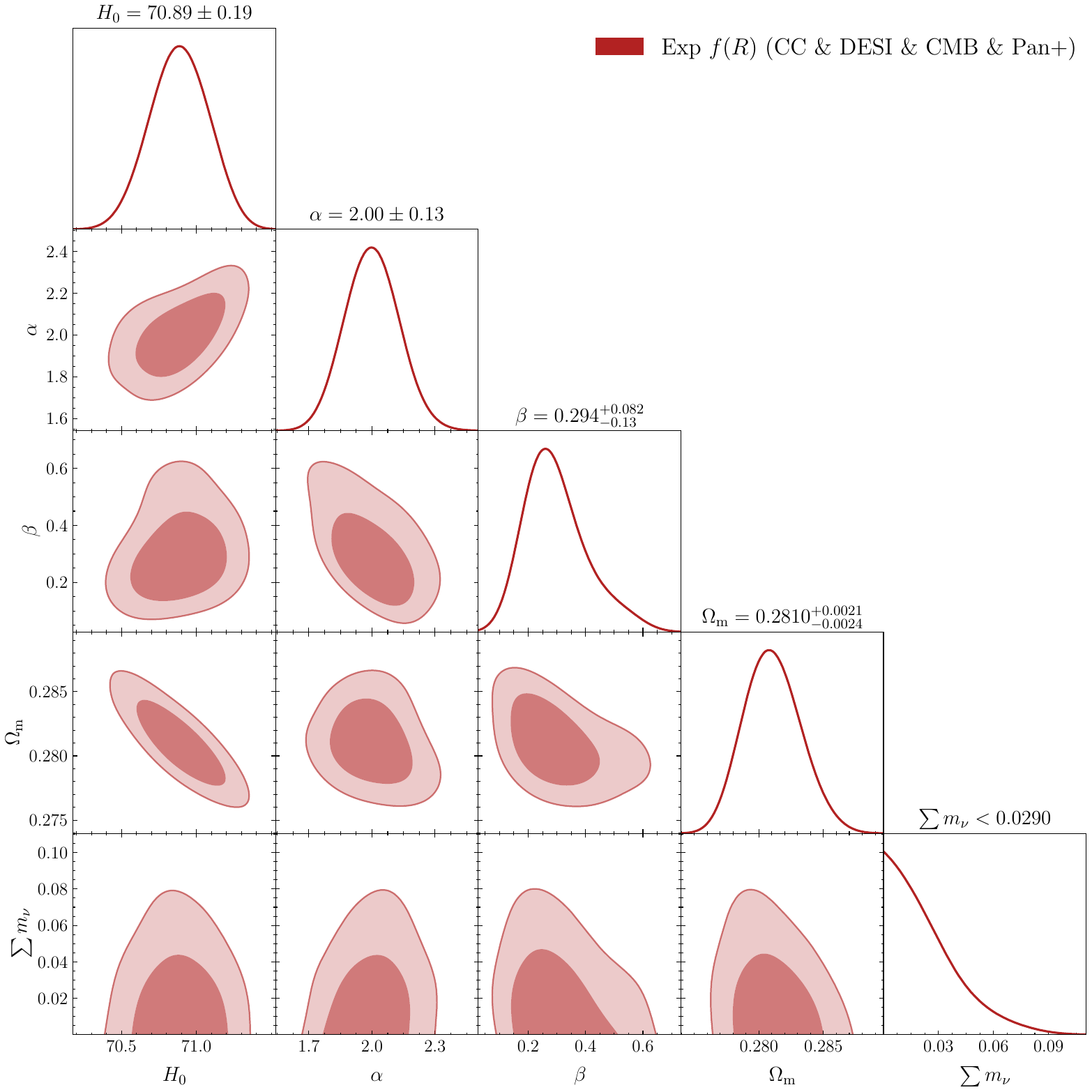}
\caption{Corner plot showing the constraints on the cosmological parameters $(H_0, \alpha,\beta,\Omegam,\sum m_\nu)$ of the exponential $f(R)$ model, obtained from the MCMC analysis using the full combination of datasets (CC, DESI DR2, CMB-$\theta_\ast$, and Pan+). The contours correspond to the $1\sigma$ ($\sim68\%$ C.L.) and $2\sigma$ ($\sim95\%$ C.L.) confidence regions, while the diagonal panels display the one-dimensional posterior distributions. After deriving the posteriors for $(H^\ast_0, \alpha,\beta,\omegab^\ast,\omegabc^\ast,\omega_\nu^\ast)$, the physical parameters of the $f(R)$ model $(H_0, \alpha,\beta,\omegab,\omegabc,\omega_\nu)$ are recovered through the mapping relations in Eq.~\eqref{eq:mapping-relations}.}
\label{fig:Corner_F_Full}
\end{center}
\end{figure}

The corner plots for the full joint analysis (all datasets combined) within the exponential $f(R)$ gravity framework are shown in Fig.~\ref{fig:Corner_F_Full}. The results demonstrate good constraining power: the contours are well closed, and also an upper limit on the total neutrino mass is obtained.
The corner plots also provide useful information on the physical interplay between the modified-gravity and cosmological sectors. The strong degeneracy between $\alpha$, $H_0$, and $\Omega_m$ indicates that modifications of the gravitational dynamics can partially compensate changes in the matter density and expansion history. The degeneracy between $\alpha$ and $\beta$ visible in Fig.~\ref{fig:Corner_F_Full} reflects the fact that similar expansion histories can be produced by different combinations of these two parameters, a feature common to models where the dark energy behavior emerges from a single exponential function of the curvature. Furthermore, the non-Gaussian posterior distribution of $\beta$ reflects the nonlinear role of the exponential correction in the gravitational Lagrangian, leading to asymmetric constraints around the best-fit value.

To verify the theoretical viability of the model across the posterior parameter space, we evaluate the conditions $f_R > 0$ and $f_{RR} > 0$, which are required for the absence of ghost and tachyonic instabilities in viable $f(R)$ theories. In particular, the condition on $f_R>0$ avoids anti-gravity effects as can be seen from Eq.~\eqref{eq:effective-Einstein-constant}. As shown in Fig.~\ref{fig:fR_stability}, both conditions are satisfied throughout the entire redshift range and within the $1\sigma$ posterior region, confirming the stability of the exponential $f(R)$ model for all constrained parameter combinations.
\\
Table~\ref{table:Best_Fit_Table} lists the best-fit values for all cosmological models and dataset combinations. Alongside the minimum of the reduced $\chi^2$, we report the Akaike Information Criterion (AIC) values \cite{Akaike:1974vps,Liddle:2007fy}. We recall that the AIC is defined as $\textrm{AIC} = \chi^2_\textrm{min} + \frac{2 n N}{N-n-1}$, where $n$ and $N$ are the numbers of free parameters in the model and data points, respectively. It provides a quantitative measure of the trade-off between model fit and complexity due to the number of extra parameters, therefore it penalizes models with many parameters. A lower AIC indicates a more statistically favored model. To compare models, we use the difference $\Delta \textrm{AIC}=\textrm{AIC}_{\Lambda\textrm{CDM}}-\textrm{AIC}_{\textrm{model}}$. By convention, $\Delta \textrm{AIC}=0$ corresponds to the reference $\Lambda$CDM model; positive differences imply that the new models behave better than $\Lambda$CDM, whereas negative differences are indicative that they do worse than $\Lambda$CDM. There is a weak evidence in favor of the new model if $0<\Delta \textrm{AIC}<2$; we say that the evidence is positive if $2<\Delta \textrm{AIC}<6$, whereas the evidence becomes strong if $6<\Delta \textrm{AIC}<10$. Finally, $\Delta \textrm{AIC}>10$ indicates a very strong evidence supporting the new model against $\Lambda$CDM. Conversely, negative values of $\Delta \textrm{AIC}$ mean that $\Lambda$CDM is supported more than the alternative model from observational datasets. We also report the Bayesian Information Criterion (BIC) values \cite{Schwarz:1978tpv}, defined as $\textrm{BIC} = \chi^2_\textrm{min} + n \ln N$, where $n$ and $N$ are again the number of free parameters and data points, respectively. Like the AIC, the BIC penalizes model complexity, but does so more severely for large datasets due to the logarithmic factor. A lower BIC indicates a more statistically favored model. Analogously to the AIC comparison, we define $\Delta \textrm{BIC} = \textrm{BIC}_{\Lambda\textrm{CDM}} - \textrm{BIC}_{\textrm{model}}$, with $\Delta \textrm{BIC} = 0$ corresponding to the $\Lambda$CDM reference. Positive values favor the alternative model over $\Lambda$CDM, while negative values indicate the opposite. Following the Jeffreys scale, the evidence against the reference model is weak if $0 < \Delta\textrm{BIC} < 2$, positive if $2 < \Delta\textrm{BIC} < 6$, strong if $6 < \Delta\textrm{BIC} < 10$, and very strong if $\Delta\textrm{BIC} > 10$.\\
When SNe Ia are excluded (upper panel of Table~\ref{table:Best_Fit_Table}), the flat $\Lambda$CDM model is statistically preferred according to the AIC. It also yields the most stringent upper bound on the total neutrino mass, \textit{i.e.} the 1$\sigma$ constraint, $\sum m_\nu <0.0536\,\textrm{eV}$, though the exponential $f(R)$ model provides nearly comparable performance, $\sum m_\nu <0.0590\,\textrm{eV}$. The inferred values of $H_0$ are consistent with the Planck results ($H_{0}=\left(67.36\pm0.54\right)\,\textrm{km s}^{-1}\,\textrm{Mpc}^{-1}$) within $1.83\,\sigma$, $1.09\,\sigma$, and $0.04\,\sigma$ for the $\Lambda$CDM, $w_0w_a$CDM, and $f(R)$ cosmologies. Notably, the uncertainties on $H_0$ are larger for the $w_0w_a$CDM, and $f(R)$ models compared to $\Lambda$CDM since the two former models are sensitive to late-time probes to infer $H_0$, and we have not included SNe Ia yet in the analysis. The best-fit values of $H_0$ are consistent in less than $1\,\sigma$ among the different cosmological scenarios, but there is a slight incompatibility in $1.6\,\sigma$ between the values inferred from $\Lambda$CDM and $w_0w_a$CDM models. Furthermore, the fitted parameters $\alpha$ and $\beta$ in the exponential $f(R)$ gravity agree well in $0.38\,\sigma$ and $0.47\,\sigma$, respectively, with those reported in \cite{Odintsov:2024woi}, $\alpha=1.002^{+0.184}_{-0.173}$ and $\beta=0.733^{+0.377}_{-0.273}$, where massive neutrinos were not included.

Including the Pan+ catalog (bottom panel of Table~\ref{table:Best_Fit_Table}) significantly affects the best-fit values. Both the $w_0w_a$CDM and exponential $f(R)$ models, being sensitive to late-time probes and deviations from $\Lambda$CDM, benefit from the addition of SNe Ia data. This fact can also be observed in Fig.~\ref{fig:Corner_Sn} for the $f(R)$ model, in which the Pan+ sample helps to break degeneracies between parameters and have more constraining power. Furthermore, the upper limits on the total neutrino mass become more stringent for all models, at 1$\sigma$ level: $\sum m_\nu <0.0126\,\textrm{eV}$ for $\Lambda$CDM, $\sum m_\nu <0.0613\,\textrm{eV}$ for $w_0w_a$CDM, and $\sum m_\nu <0.0290\,\textrm{eV}$ for the $f(R)$ cosmology. We note that these bounds are quoted at $1\sigma$ ($68\%$ C.L.), whereas cosmological neutrino mass constraints are more commonly reported at $95\%$ C.L. A direct comparison with the oscillation lower bound $\sum m_\nu \gtrsim 0.06\,\mathrm{eV}$ should therefore be made at the same confidence level. While the exponential $f(R)$ model yields a less stringent upper limit on $\sum m_\nu$ compared to $\Lambda$CDM, the tension with the oscillation lower bound is not fully resolved, and our conclusions on this point should be interpreted as a mild relaxation rather than a resolution of the neutrino-mass tension.

The inclusion of these data leads to higher $H_0$ values, even when combined with DESI DR2 datasets and CMB likelihood. These new values of $H_0$ are mutually consistent in $1\,\sigma$ for the proposed cosmological scenarios, but there is a discrepancy in $2.3\,\sigma$ between the values obtained within $\Lambda$CDM and $f(R)$ models. The relative uncertainty on $H_0$ decreases by $\sim65\%$, $\sim54\%$, and $\sim86\%$ for the $\Lambda$CDM, $w_0w_a$CDM, and $f(R)$ cosmologies, respectively, compared to the corresponding cases without SNe Ia. Although one combination of datasets is included in the other one, to quantify the impact of SNe Ia on the value of $H_0$, we compare the best-fit values obtained with and without the Pan+ sample. We notice a large tension of $7.5\,\sigma$ between the best-fit values of $H_0$ within the $\Lambda$CDM model with and without SNe Ia. This tension is significantly reduced at the level of $2.4\,\sigma$ and $2.8\,\sigma$ for the $w_0w_a$CDM and $f(R)$ models, respectively. 

Regarding the values of the extra parameters, $\alpha$ and $\beta$ now are consistent in $2.0\,\sigma$ and $0.9\,\sigma$, respectively, with the previous results without SNe Ia, while the level of compatibility decreases at $4.4\,\sigma$ and $1.5\,\sigma$, respectively, if compared to the results presented in \cite{Odintsov:2024woi} due to the impact of massive neutrinos in the late-time observational datasets. 
Interestingly, with the inclusion of the Pan+ sample, both the $w_0w_a$CDM and $f(R)$ cosmologies are statistically preferred over $\Lambda$CDM according to the AIC, with $\Delta\textrm{AIC}=43.75$ and $\Delta\textrm{AIC}=20.46$, respectively, both indicating very strong evidence in favor of the alternative models. The BIC corroborates this conclusion: $\Delta\textrm{BIC}=32.84$ for $w_0w_a$CDM and $\Delta\textrm{BIC}=9.55$ for the $f(R)$ model, corresponding to very strong and strong evidence, respectively, against $\Lambda$CDM. Conversely, without Pan+, both criteria yield negative differences for the alternative models, indicating that $\Lambda$CDM is preferred by the CC\,\&\,DESI\,\&\,CMB dataset alone.

\begin{figure}[htb!]
\begin{flushleft}
\includegraphics[width=1\linewidth]{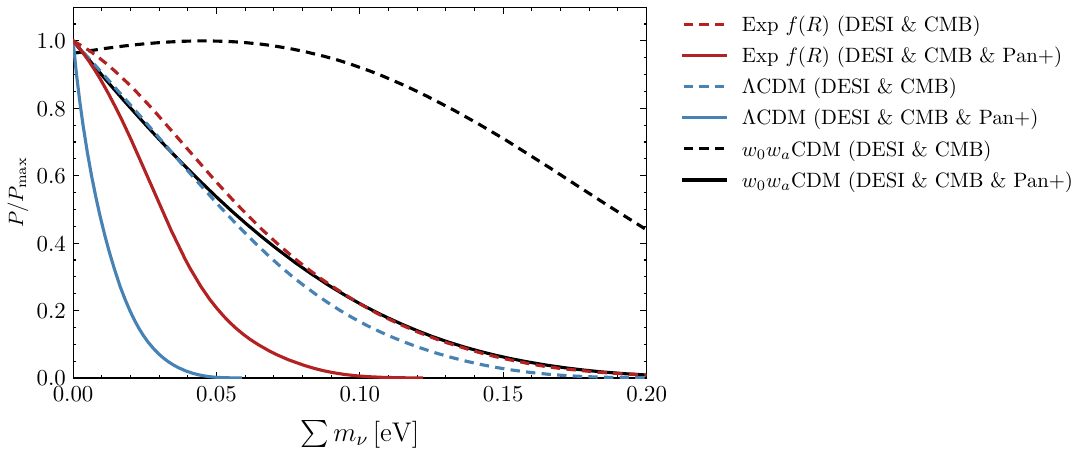}
\caption{Marginalized 1$\sigma$ posterior distribution for the parameter $\sum m_\nu$ for different cosmological models and combination of datasets. Red, blue, and black curves are referred to the exponential $f(R)$ gravity, $\Lambda$CDM, and $w_0w_a$CDM models, respectively. A dashed curve is obtained from the analysis using the combination of CC, DESI, and CMB datasets, while a continuous curve is relative to the same datasets but including the Pan+ sample.}
\label{fig:SumNu_Comparison_CPL}
\end{flushleft}
\end{figure}

In Fig.~\ref{fig:SumNu_Comparison_CPL} we compare the marginalized posterior distribution for the $\sum m_\nu$ within the three cosmological scenarios, considering different combinations of datasets with and without SNe Ia. We note that the $\Lambda$CDM model is characterized by tighter bounds on $\sum m_\nu$ due to the fact that the other two cosmological models are characterized by a large number of parameters, but the exponential $f(R)$ cosmology can constrain better $\sum m_\nu$ than the $w_0w_a$CDM model. Moreover, from Fig.~\ref{fig:SumNu_Comparison_CPL}, we can also see that the inclusion of the SNe Ia catalog helps to have tighter constraints on $\sum m_\nu$. Focusing on the exponential $f(R)$ cosmology, we see in Fig.~\ref{fig:Corner_nu} that we have similar constraints on the cosmological parameters with and without massive neutrinos, except for the CDM component that changes its central value due to the transition from non-relativistic to relativistic particles.
Finally, Figs.~\ref{fig:Corner_w0wa} summarize the constraints on cosmological parameters for the three models, considering the datasets without SNe Ia, while Fig.~\ref{fig:Corner_w0wa_Sn} is analogous with the inclusion of SNe Ia. Comparing these two figures, we note that $f(R)$ gravity benefits from the addition of SNe Ia, as already stressed, reducing the uncertainties of the cosmological parameters. This fact also implies a noticeable departure from the $\Lambda$CDM model, highlighting the role of low-redshift datasets to discriminate between alternative models and $\Lambda$CDM.  

\section{Conclusions} \label{sec:conclusions}
In this work, we have performed a comprehensive MCMC analysis to constrain the parameters of the exponential $f(R)$ gravity model in the presence of massive neutrinos. Among the various modified gravity scenarios, $f(R)$ theories represent one of the most theoretically motivated frameworks, as they naturally introduce additional scalar degrees of freedom that can modify cosmic dynamics while remaining consistent with GR and $\Lambda$CDM cosmology at high redshifts. The exponential realization of $f(R)$ gravity, investigated here, provides a smooth transition from the standard GR regime in the early Universe to a modified phase at late times, where deviations from $\Lambda$CDM may become observationally relevant.\\
Our analysis confirms that this class of models can reproduce the main features of the observed cosmic expansion while introducing an effective dynamical dark energy behavior capable of partially alleviating the Hubble tension (see the discussion in Sec.~\ref{sec:constraints}). The inclusion of massive neutrinos, whose total mass is bounded by terrestrial oscillation experiments to $\sum m_\nu \gtrsim 0.06,\text{eV}$, plays a key role in shaping cosmological observables. Within the $\Lambda$CDM framework, cosmological data tend to prefer values of $\sum m_\nu$ below this threshold, leading to a mild but persistent tension between cosmological and laboratory constraints. As recently highlighted by the DESI Collaboration, models with a dynamical dark energy component, such as the CPL parameterization, can relax this tension by broadening the allowed parameter space for $\sum m_\nu$. Since the $w_0w_a$CDM model misses a theoretical interpretation, we tried to see if the exponential $f(R)$ model could lead to the same result but with its robustness.\\
In our exponential $f(R)$ analysis, we find that this modified gravity scenario leads to a mild relaxation of the neutrino mass upper bound compared to $\Lambda$CDM, although the resulting $1\sigma$ constraint $\sum m_\nu < 0.0290\,\mathrm{eV}$ remains below the oscillation lower limit of $\sim 0.06\,\mathrm{eV}$, and the tension is therefore not fully resolved. The constraints on $\sum m_\nu$ also remain tighter than those obtained with the CPL model (Fig. \ref{fig:SumNu_Comparison_CPL}). Nonetheless, dynamical dark energy or the inclusion of a dynamical degree of freedom in the gravitational sector leads to a statistically improved fit to the combined dataset (CC, DESI DR2, CMB-$\theta_\ast$, and Pan+), compared to the $\Lambda$CDM scenario, as reflected by the AIC and BIC results (see Table~\ref{table:Best_Fit_Table}).\\
Overall, the exponential $f(R)$ gravity with massive neutrinos represents a theoretically robust and observationally viable extension of the standard cosmological model. While it provides only a moderate alleviation of current cosmological tensions, it offers a consistent framework in which both gravitational modifications and neutrino physics can interplay to shape late-time cosmology. We emphasize that the present analysis is restricted to background cosmology. Since both the scalar degree of freedom of $f(R)$ gravity and massive neutrinos affect the growth of matter perturbations, future investigations including redshift-space distortions, weak lensing, galaxy clustering, and $f\sigma_8$ measurements will be necessary to fully assess the viability of the model. Future work should also explore broader classes of $f(R)$ models to identify those capable of reconciling cosmological and terrestrial neutrino mass constraints more effectively. Extending this analysis to other modified gravity scenarios, such as scalar-tensor or Gauss–Bonnet theories, and incorporating alternative SNe Ia datasets from DES-Y5, and forthcoming datasets from DESI or next-generation CMB surveys will further test the predictive power and parameter space of these models to alleviate cosmological tensions.

\section*{Acknowledgments}
We are very thankful to Sandeep Haridasu and Andrea Cozzumbo for useful discussions. SD is funded by MCIN/AEI/10.13039/501100011033 and FSE+, reference PRE2021-098098 and the program Unidad de Excelencia María de Maeztu
CEX2020-001058-M. TS is supported by the Della Riccia foundation grant 2025.

\begin{widetext}

\begin{figure*}[htb]
\begin{figure}[H]
    \centering
    \begin{minipage}{0.49\textwidth}
        \centering
    \centering
    \includegraphics[width=1\linewidth]{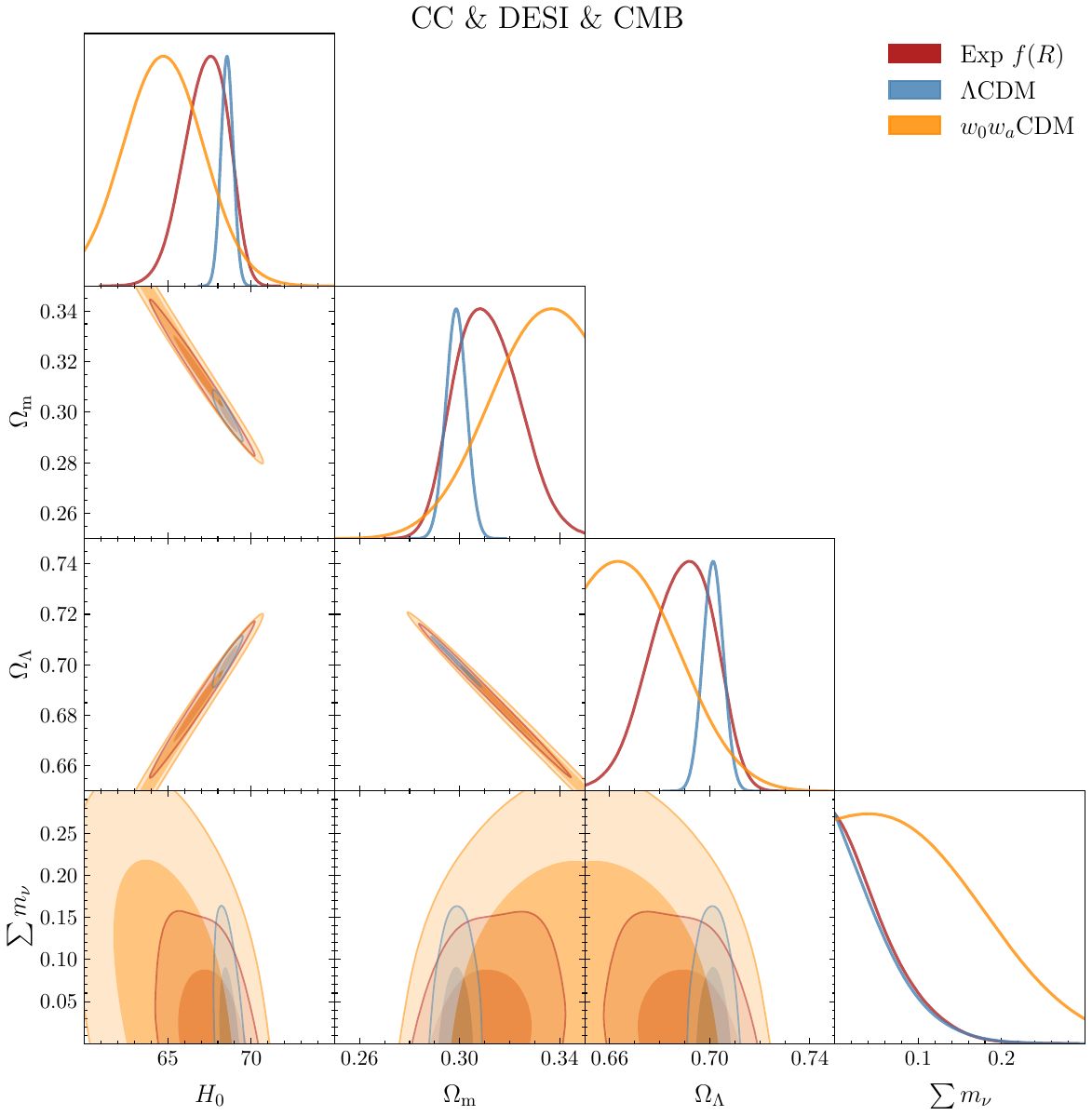}
    \end{minipage}%
    \hfill
    \raisebox{0ex}{\begin{minipage}[c]{0.49\textwidth}
        \centering
    \includegraphics[width=1\linewidth]{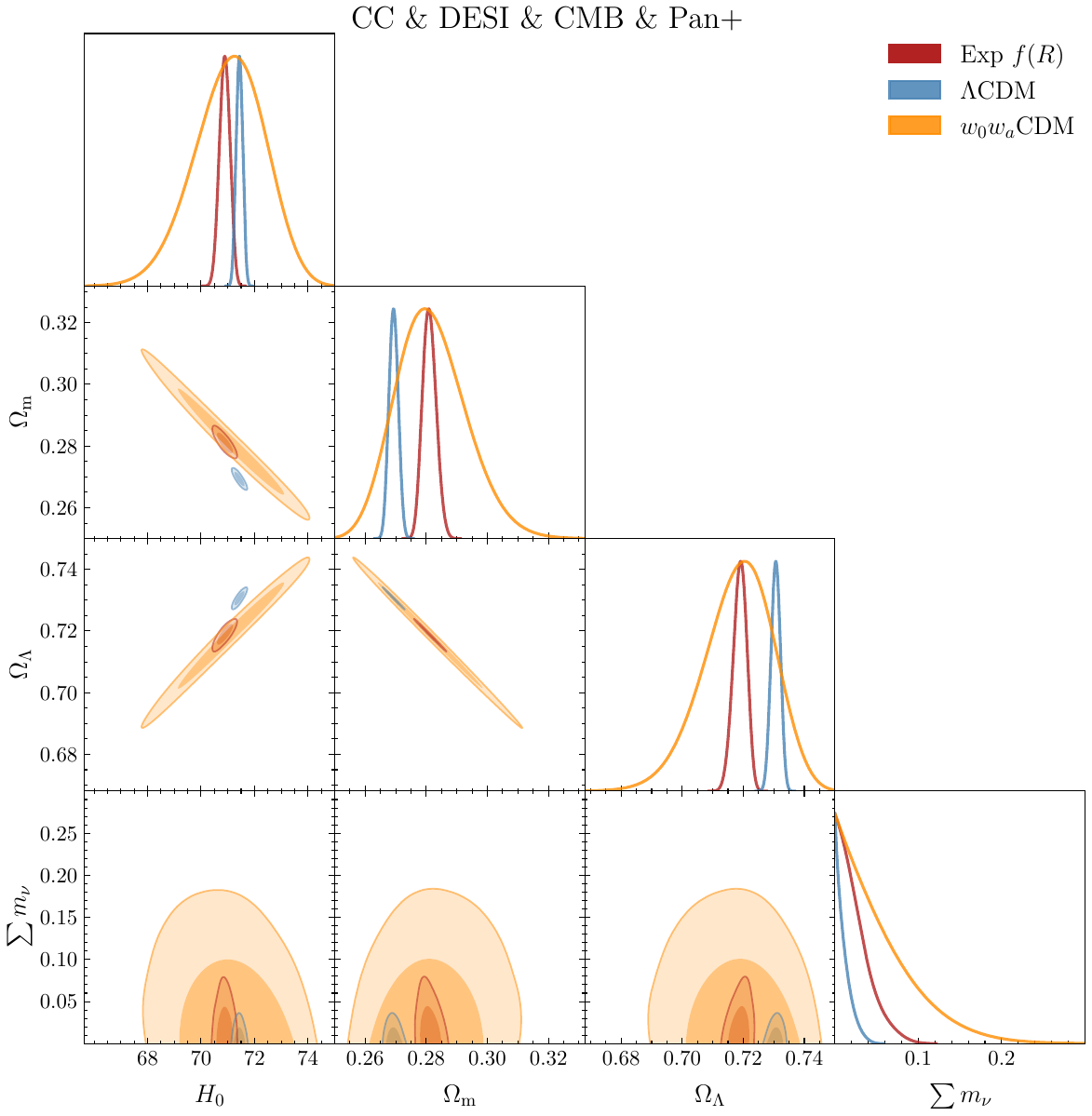}
    \end{minipage}}%
    \par
    \begin{minipage}[t]{0.49\textwidth}
        \centering
    \caption{Corner plot for the parameters $(H_0,\Omegam,\Omega_\Lambda,\sum m_\nu)$ in the joint datasets without the SNe catalog.}
    \label{fig:Corner_w0wa}
    \end{minipage}%
    \hfill
    \begin{minipage}[t]{0.49\textwidth}
        \centering
    \caption{Corner plot for the parameters $(H_0,\Omegam,\Omega_\Lambda,\sum m_\nu)$ in the joint datasets with the SNe catalog.}
    \label{fig:Corner_w0wa_Sn}
    \end{minipage}%
\end{figure}
\end{figure*}

\begin{figure*}[htb]
\begin{figure}[H]
    \centering
    \begin{minipage}{0.49\textwidth}
        \centering
    \centering
        \includegraphics[width=1\linewidth]{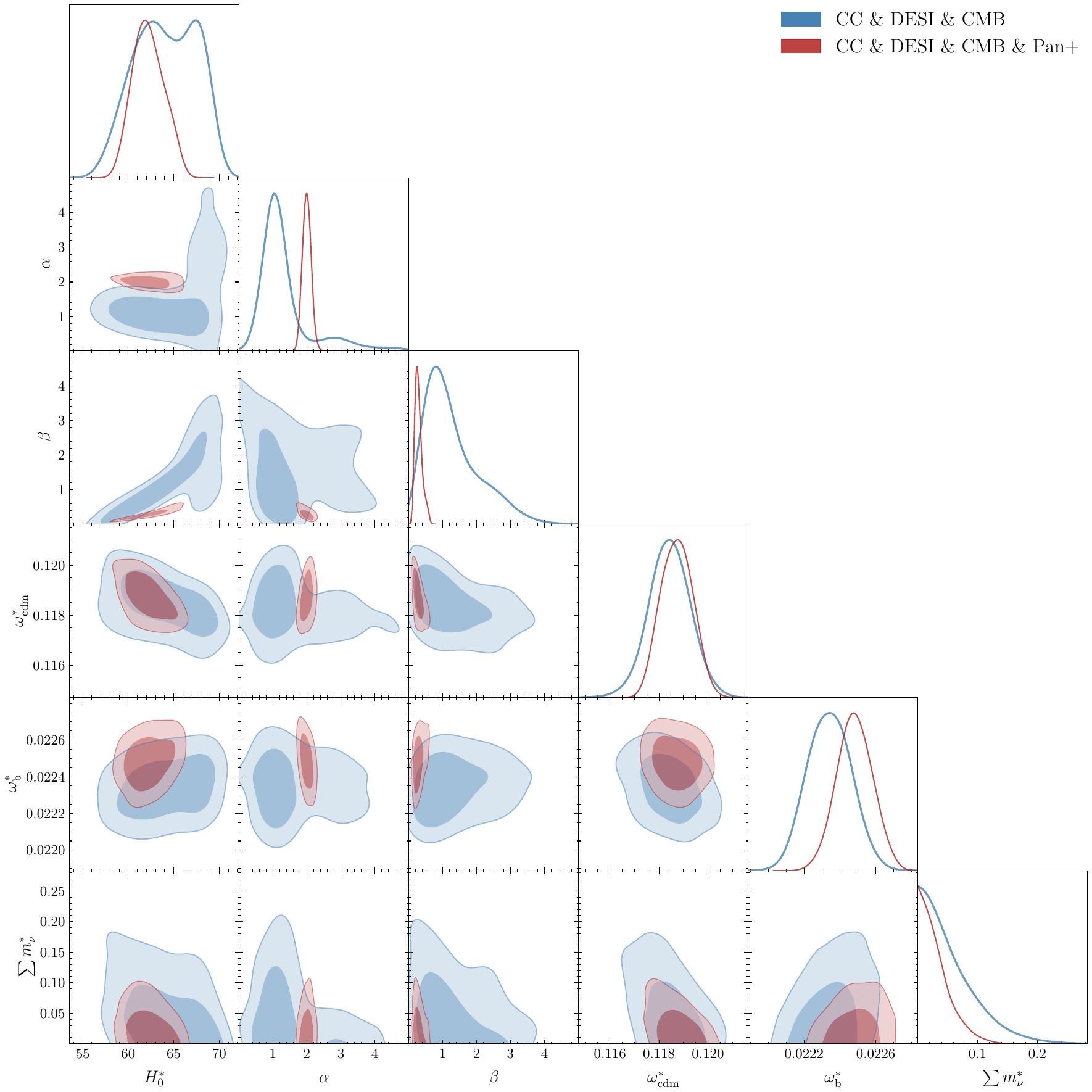}
    \end{minipage}%
    \hfill
    \raisebox{0ex}{\begin{minipage}[c]{0.49\textwidth}
        \centering
        \includegraphics[width=1\linewidth]{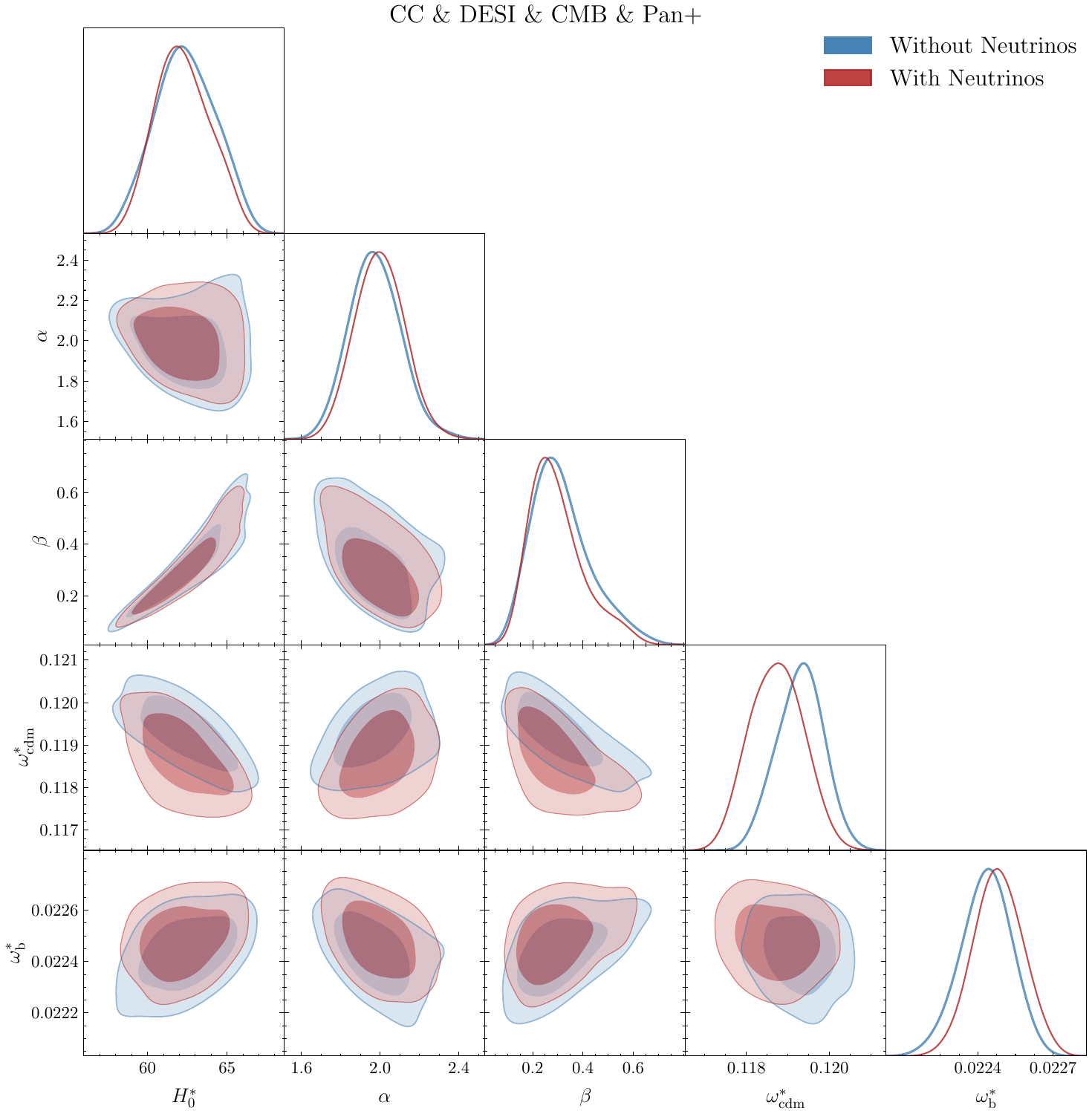}
    \end{minipage}}%
    \par
    \begin{minipage}[t]{0.49\textwidth}
        \centering
    \caption{Comparison between the corner plots for the parameters $(H_0^*,\alpha,\beta,\omegab^*,\omega_\text{cdm}^*,\sum m_\nu^*)$ in the case with and without the SNe catalog including neutrinos. With an abuse of notation, we refer to $\sum m_\nu^*$ as $\sum m_\nu^*=E(0)^2\sum m_\nu$.}
    \label{fig:Corner_Sn}
    \end{minipage}%
    \hfill
    \begin{minipage}[t]{0.49\textwidth}
        \centering
    \caption{Comparison between the corner plots for the parameters $(H_0^*,\alpha,\beta,\omegab^*,\omega_\text{cdm}^*)$ in the case with and without neutrinos.}
    \label{fig:Corner_nu}
    \end{minipage}%
\end{figure}
\end{figure*}

\begin{figure*}[htb]
\begin{figure}[H]
    \centering
    \begin{minipage}[c]{1.\textwidth}
        \centering
    \centering
    \includegraphics[width=1\linewidth]{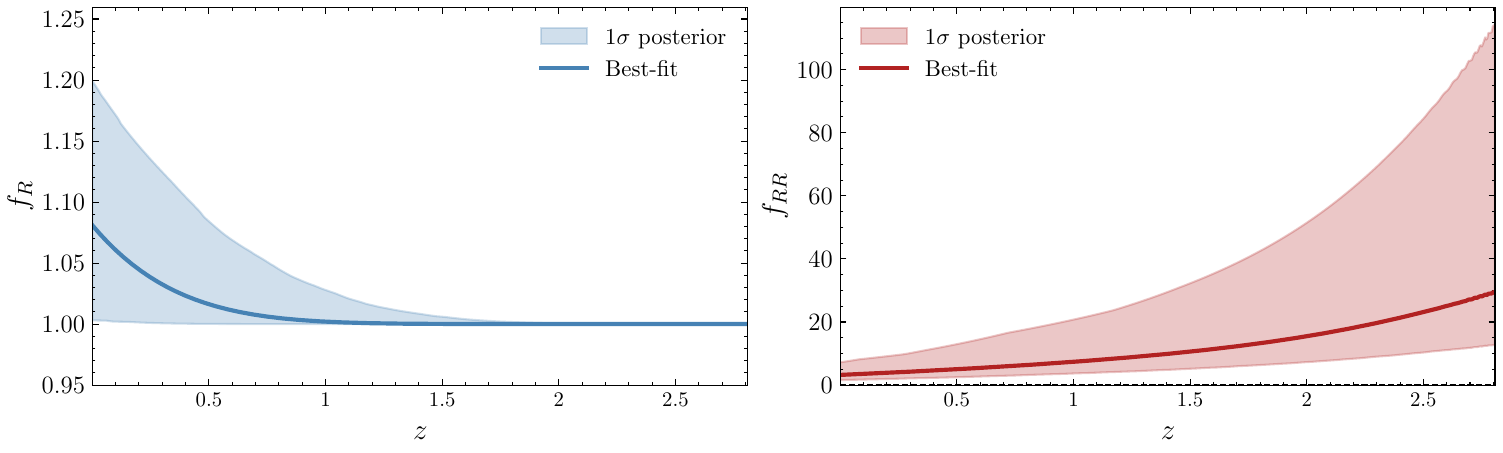}
        \centering
    \caption{Posterior evaluation of the viability conditions $f_R > 0$ and $f_{RR} > 0$ for the exponential $f(R)$ model, using the CC\,\&\,DESI\,\&\,CMB\,\&\,Pan+ dataset combination. Both quantities are shown as a function of redshift, with the shaded band representing the $1\sigma$ posterior region and the solid line indicating the best-fit solution. The conditions $f_R > 0$ and $f_{RR} > 0$ are satisfied across the entire redshift range and throughout the posterior parameter space, confirming the theoretical viability of the model.}
    \label{fig:fR_stability}
    \end{minipage}
\end{figure}
\end{figure*}
\end{widetext}

\setlength{\arrayrulewidth}{0.1mm}
\setlength{\tabcolsep}{4pt}
\renewcommand{\arraystretch}{1.2}
\begin{figure}
    \centering
    \begin{minipage}{0.49\textwidth}
        \centering
    \centering
    \begin{tabular}{|c | c | c | c | c | c | c | c |} 
 \hline
 \multicolumn{8}{|c|}{CC} \\
 \hline \hline
 $z$ & $H(z)$ & $\sigma_H$ & Refs & $z$ & $H(z)$ & $\sigma_H$ & Refs \\ 
\hline
$0.070$ & $69$ & $19.6$ & \cite{zhang2014four} & $0.4783$ & $80.9$ & $9$ &  \cite{Moresco_2016} \\
$0.090$ & $69$ & $12$ & \cite{Simon:2004tf} & $0.480$ & $97$ & $62$ & \cite{Daniel_Stern_2010}\\
$0.120$ & $68.6$ & $26.2$ & \cite{zhang2014four} & $0.593$ & $104$ & $13$ & \cite{M_Moresco_2012}\\
$0.170$ & $83$ & $8$ &  \cite{Simon:2004tf} & $0.6797$ & $92$ & $8$ & \cite{M_Moresco_2012}\\
$0.1791$ & $75$ & $4$ & \cite{M_Moresco_2012} & $0.7812$ & $105$ & $12$ & \cite{M_Moresco_2012}\\
$0.1993$ & $75$ & $5$ & \cite{M_Moresco_2012} & $0.8754$ & $125$ & $17$ & \cite{M_Moresco_2012}\\
$0.200$ & $72.9$ & $29.6$ & \cite{zhang2014four} & $0.880$ & $90$ & $40$ & \cite{Daniel_Stern_2010}\\
$0.270$ & $77$ & $14$ & \cite{Simon:2004tf}  & $0.900$ & $117$ & $23$ &  \cite{Simon:2004tf} \\
$0.280$ & $88.8$ & $36.6$ & \cite{zhang2014four} & $1.037$ & $154$ & $20$ & \cite{M_Moresco_2012}\\
$0.3519$ & $83$ & $14$ & \cite{M_Moresco_2012} & $1.300$ & $168$ & $17$ &  \cite{Simon:2004tf}\\
$0.3802$ & $83$ & $13.5$ & \cite{Moresco_2016} & $1.363$ & $160$ & $33.6$ &\cite{Moresco_2015} \\
$0.400$ & $95$ & $17$ &  \cite{Simon:2004tf}  & $1.430$ & $177$ & $18$ &   \cite{Simon:2004tf}\\
$0.4004$ & $77$ & $10.2$ &  \cite{Moresco_2016} & $1.530$ & $140$ & $14$ & \cite{M_Moresco_2012}\\
$0.4247$ & $87.1$ & $11.2$ &  \cite{Moresco_2016} & $1.750$ & $202$ & $40$ & \cite{M_Moresco_2012}\\
$0.4497$ & $92.8$ & $12.9$ &  \cite{M_Moresco_2012} & $1.965$ & $186.5$ & $50.4$ & \cite{Moresco_2015}\\
$0.470$ & $89$ & $34$ &  \cite{Ratsimbazafy_2017} &   &   &   & \\
\hline
\end{tabular} 
\caption{Values of $H(z)$ at different redshift and its uncertainty obtained for the CC analysis.}
    \label{table:CC_Data}
    \end{minipage}%
\end{figure}

\phantomsection


\bibliographystyle{apsrev4-2}

\bibliography{Bibliography}

\end{document}